\documentclass[aps,prl,twocolumn,10pt,amsmath,superscriptaddress,amssymb,longbibliography]{revtex4}%"longbibliography" will display full title and authors' name
\usepackage[normalem]{ulem}
\usepackage{epsfig}
\usepackage{epstopdf}
\usepackage{dcolumn}
\usepackage{amsbsy}
\usepackage{bm}
\usepackage{color,CJK}
\usepackage{amsmath,amssymb,amsfonts}
\usepackage{appendix}
\usepackage{graphicx}
\usepackage{algorithm}
\usepackage{enumerate}
\usepackage{makeidx}
\usepackage{braket}
\usepackage[usenames,dvipsnames]{xcolor}
\usepackage[driverfallback=dvipdfmx,colorlinks,linkcolor=blue,citecolor=blue,urlcolor=blue]{hyperref}

\begin{document}

\title{Atomic Bose-Einstein condensate in twisted-bilayer optical lattices}

\author{Zengming Meng}
\email[These authors contributed equally to this work.]{}

\author{Liangwei Wang}

\email[These authors contributed equally to this work.]{}

\author{Wei Han}

\author{Fangde Liu}

\author{Kai Wen}
\affiliation{State Key Laboratory of Quantum Optics and Quantum
Optics Devices, \\  Institute of Opto-Electronics, \\ Collaborative Innovation Center of Extreme Optics, Shanxi
University, Taiyuan, Shanxi 030006, People's Republic of China }

\author{Chao Gao}
\affiliation{Department of Physics, Zhejiang Normal University, Jinhua, 321004, People's Republic of China }

\author{Pengjun Wang}

\affiliation{State Key Laboratory of Quantum Optics and Quantum
Optics Devices, \\  Institute of Opto-Electronics, \\ Collaborative Innovation Center of Extreme Optics, Shanxi
University, Taiyuan, Shanxi 030006, People's Republic of China }

\author{Cheng Chin}
\affiliation{James Franck Institute, Enrico Fermi Institute, Department of Physics,
University of Chicago, Chicago, Illinois 60637, USA}

\author{Jing Zhang}
\email[Corresponding author email: ]{jzhang74@sxu.edu.cn}

\affiliation{State Key Laboratory of Quantum Optics and Quantum
Optics Devices, \\  Institute of Opto-Electronics, \\ Collaborative Innovation Center of Extreme Optics, Shanxi
University, Taiyuan, Shanxi 030006, People's Republic of China }

\affiliation{Hefei National Laboratory, Hefei, Anhui 230088, P. R. China}

\date{\today }

\begin{abstract}
Observation of strong correlations and superconductivity in twisted-bilayer-graphene~\cite{Cao2018-1,Cao2018-2,Yankowitz1059,Lu2019} has stimulated tremendous interest in fundamental and applied physics~\cite{WANG2019100099,Andrei2020,Balents2020,Kennes2021}. In this system, the superposition of two twisted honeycomb lattices, generating a moir$\acute{\mathrm{e}}$ pattern, is the key to the observed flat electronic bands, slow electron velocity and large density of states~\cite{PhysRevLett.99.256802,PhysRevB.81.161405,Bistritzer12233,PhysRevB.85.195458}. Extension of the twisted-bilayer system to new configurations is highly desired, which can open up exciting prospects to investigate twistronics beyond bilayer graphene. Here we demonstrate a quantum simulation of superfluid-to-Mott insulator transition in twisted-bilayer square lattices based on atomic Bose-Einstein condensates loaded into spin-dependent optical lattices. The lattices are made of two sets of laser beams that independently address atoms in different spin states, which form the synthetic dimension accommodating the two layers. The interlayer coupling is highly controllable by microwave field, which enables the occurrence of flat lowest band and novel correlated phases in the strong coupling limit. We directly observe the spatial moir$\acute{\mathrm{e}}$
pattern and the momentum diffraction, which confirm the presence of two forms of superfluid and a modified superfluid to insulator transition in twisted-bilayer lattices. Our scheme is generic and can be applied to different lattice geometries and for both boson and fermion systems. This opens up a new direction for exploring moir$\acute{\mathrm{e}}$ physics in ultracold atoms with highly controllable optical lattices.
\end{abstract}

\maketitle

Novel band structures in lattice systems often lead to new material functions and discoveries. Twistronics, originating from the twisted-bilayer-graphene as a tunable experimental platform~\cite{Cao2018-1,Cao2018-2,Yankowitz1059,Lu2019,WANG2019100099,Andrei2020,Balents2020,Kennes2021} has attracted broad attention in recent years and launched intensive theoretical research. Here overlaying two graphene layers with a small relative angle exhibit the rich phase diagram, such as the coexistence of unconventional superconductivity and correlated insulating phases  ~\cite{Cao2018-2,Yankowitz1059,Lu2019}. In recent years, many examples of twisted-bilayer are discovered with remarkable physical properties not present in their untwisted counterparts. Recently,
photonic moir$\acute{\mathrm{e}}$ lattices are explored for their capabilities in localizing and delocalizing light~\cite{Wang2020,Huang2016,Fu2020} and twisted-bilayer materials for engineering the photonic dispersion~\cite{Hu2020}.

% Ultracold atomic gases provide a versatile platform for exploring many interesting quantum phenomena~\cite{RMP1,RMP2,RMP3}, which give insights into systems that are difficult to realize in solid state systems. In particular,

Ultracold atoms in optical lattices constitute an ideal platform to simulate emerging many-body phenomena in condensed matter physics~\cite{RevModPhys.80.885,Lewenstein-book,Windpassinger_2013}. Different optical lattice geometries can be realized by interfering different sets of laser beams ~\cite{Soltan-Panahi2011,Wirth2011,Tarruell2012,PhysRevLett.108.045305,Taiee1500854,Gall2021}.
% 2D quasicrystal potential formed by superimposing four independent 1D lattice for single component of BEC is investigated~\cite{Viebahn2019} and localization in it is observed~\cite{Sbroscia2020}.
In particular, a scheme of simulating twisted-bilayer lattice is recently proposed using two overlapping optical lattices~\cite{PhysRevA.100.053604,PhysRevLett.126.103201}. Other schemes for simulating bilayer heterostructures have also been put forward~\cite{Gra__2016,PhysRevLett.125.030504}. These schemes are based on coherent coupling between spin states of atoms, which simulates interlayer tunnelling along an artificial, synthetic dimension ~\cite{PhysRevLett.108.133001,PhysRevLett.112.043001,Ozawa2019}.

% Although simulating twisted bilayer with ultracold atoms shows the attractive prospects~\cite{Chuanwei2020}, these proposals have not been realized in laboratories.

\begin{figure}
\includegraphics[width=3.5in]{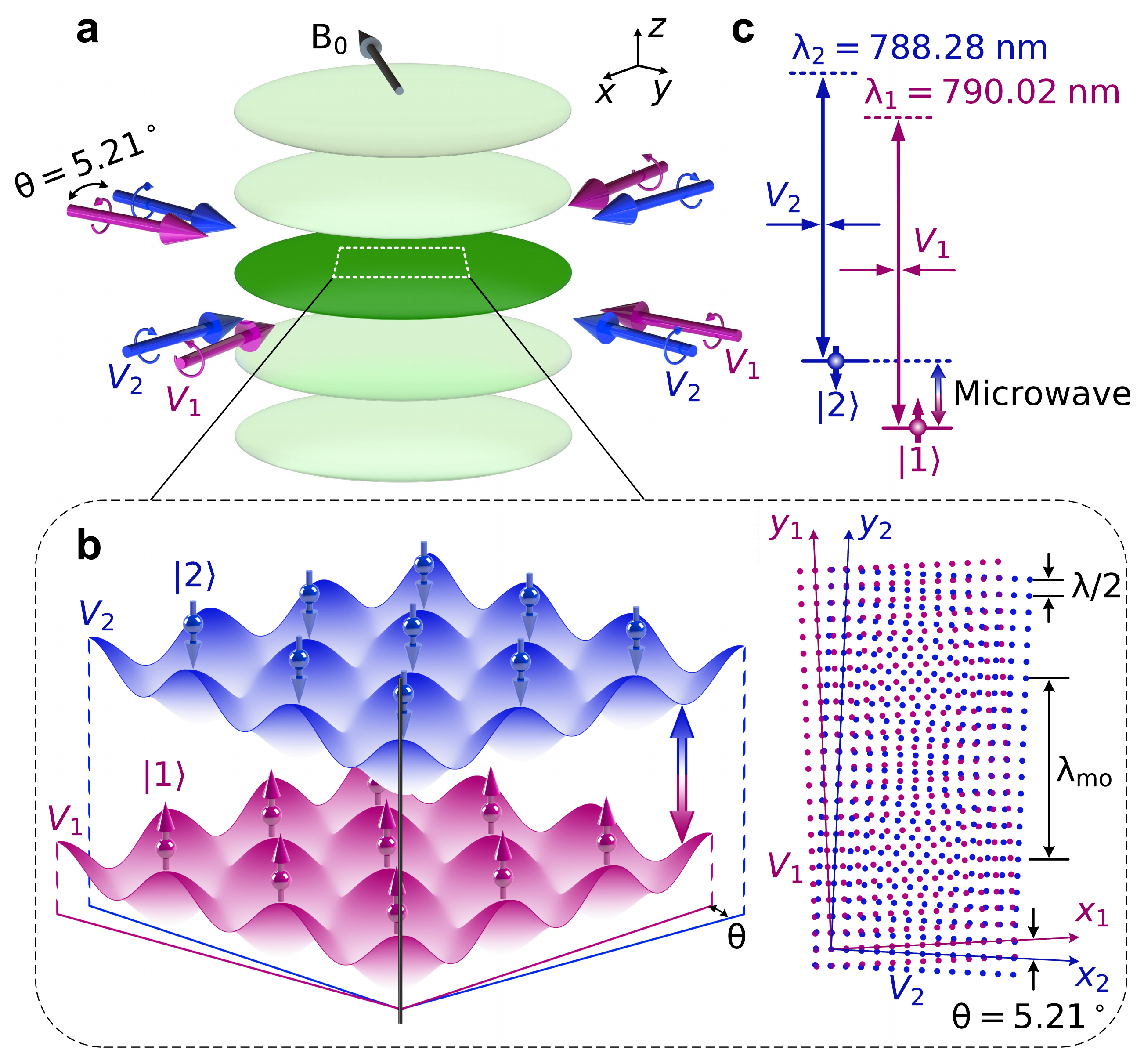}
%\begin{figure*}
%\centerline{
%\includegraphics[width=0.45\linewidth]{Fig1.pdf}
%} \vspace{0.1in}
%\setlength{\columnwidth}{3.2in}
%\centerline{
\caption{ \textbf{Simulation of twisted-bilayer systems based on atoms in spin-dependent optical lattices.} $\bf{a}$~Atoms are loaded into a single layer, 2D pancake-like potential formed by a vertical optical lattice (green) in the $z$-direction. Two sets of square optical lattices $V_1$ (purple) and $V_2$ (blue) on the horizontal plane with a small relative angle $\theta=5.21^{\circ}$ form a spin-dependent lattice potential and confine Rb atoms in spin state $\left\vert 1 \right\rangle$ (up-arrows) and $\left\vert 2 \right\rangle$ (down-arrows) independently. A magnetic field is applied in the $x$-$y$ plane along the 45$^{\circ}$ diagonal of the $V_2$ lattice. The lattice beams for $V_1$ and $V_2$ are set with opposite circular polarization to generate the vector shift with the opposite sign.  $\bf{b}$~Left panel: Sketch of the bilayer lattices in the synthetic dimension. The interlayer tunnelling is controlled by an MW field. Right panel: Superimposed lattice structure with the lattice constant $\lambda/2$ and much larger moir$\acute{\mathrm{e}}$ length $\lambda_{mo}$. $\bf{c}$~Energy diagram of the two ground Zeeman states $\left\vert 1 \right\rangle$ and $\left\vert 2 \right\rangle$ and the associated lattice beams at the tune-out wavelengths $\lambda_1=790.02$~nm and $\lambda_2=788.28$~nm.}
\label{Fig1}
%}
%\end{figure*}
\end{figure}

In this paper we demonstrate Bose-Einstein condensates (BEC) of Rubidium-87 ($^{87}$Rb) atoms loaded into a pair of twisted-bilayer optical lattices. Two overlapping lattices $V_1$ and $V_2$ are formed by interfering laser beams at the specific ``tune-out'' wavelengths ~\cite{leblanc2007species,arora2011tune,Wen2021} $\lambda_1$ and $\lambda_2$ with proper polarizations such that atoms in spin state $\left\vert 1\right\rangle\equiv\left\vert F=1, m_F=1\right\rangle$ and state $\left\vert 2\right\rangle\equiv \left\vert F=2, m_F=0\right\rangle$ only experience the lattice potential $V_1$ and $V_2$, respectively, see Fig.~1. Here $F$ and $m_F$ are the angular momentum and projection quantum numbers in the $^{87}$Rb ground state manifold. Each set of the laser beams forms a two-dimensional (2D) square lattice on the horizontal x-y plane and the twist of the two lattices is realized by orienting the beams of different wavelengths with a small relative angle $\theta=5.21^{\circ}$. The sample is tightly confined in the vertical $z$-direction such that the sample is in the quasi-2D regime (see Supplementary Material section I for details).

% using spin-dependent optical lattice with square geometries for $^{87}Rb$ Bose-Einstein condensate. First, we load a 3D shaped BEC into a single layer of the 2D pancakes at maximum lattice spacing for accordion lattice and then compress the pancake adiabatically to reach a deep 2D regime by reducing the lattice spacing. We employ tune-out wavelength for spin-dependent optical lattice, in which ac Stark shift cancels. Two internal spin states have the different tune-out wavelengths when considering the contributions of both the scalar and vector shift. Therefore, we can generate two spin-dependent square optical lattices with a twisted angle between them, which trap independently two ground spin states.

%. We can drive the transition between two internal spin states by micro-wave (MW) field to obtain the intralayer interaction, which leads to the synthetic layer dimension. Therefore, the inter- and intralayer coupling can be controlled independently in experiment.

\begin{figure}
\includegraphics[width=3.5in]{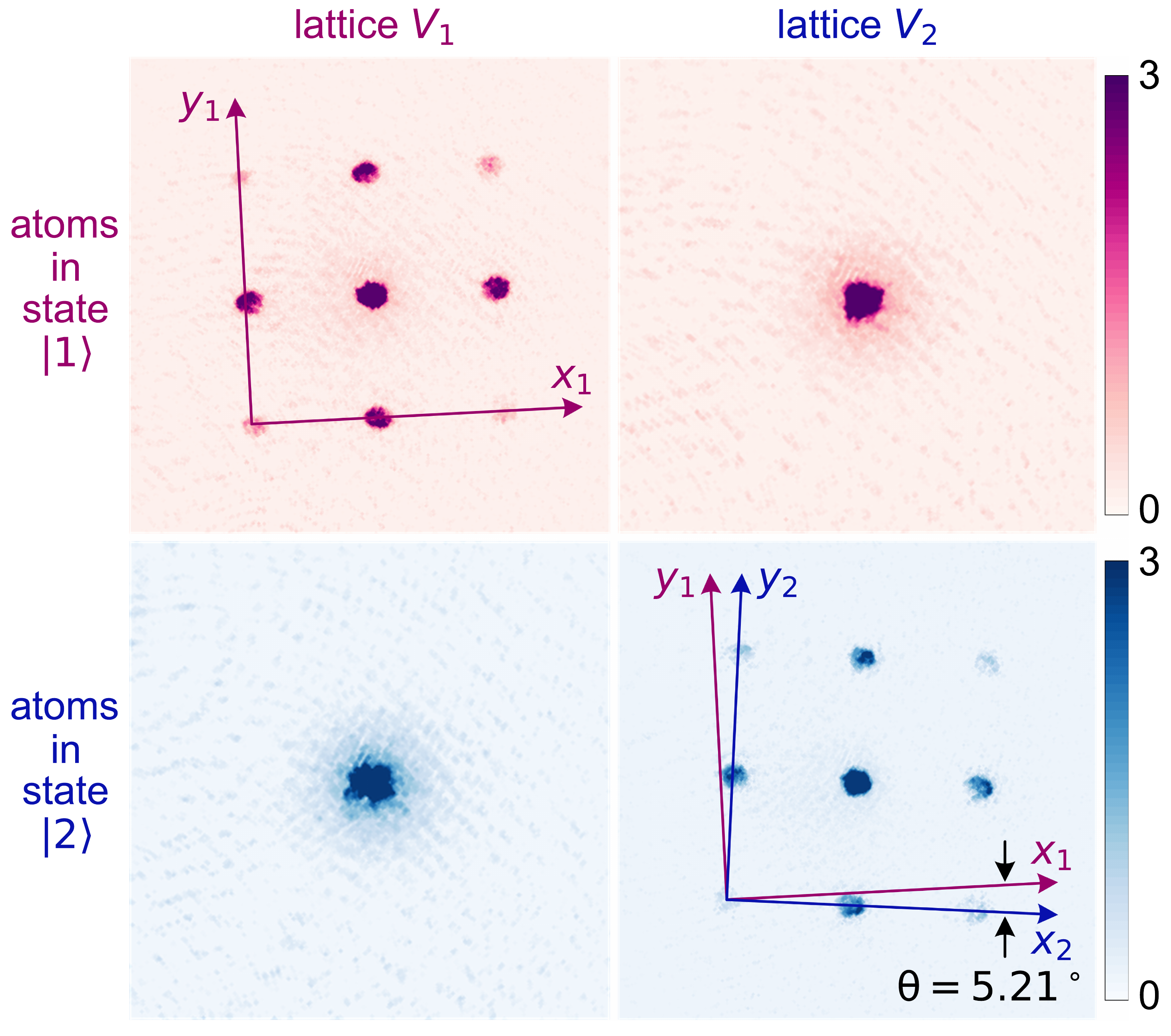}
%\setlength{\columnwidth}{3.2in}
%\centerline{
\caption{ \textbf{Independent diffraction of atoms in different spin states by the twisted-bilayer optical lattices.} The optical lattice potential is applied to the atomic BEC with a short duration of 4~$\mu$s. The images show diffraction patterns of the atoms after 18~ms free space expansion. At the tune-out wavelength $\lambda_1=790.02$~nm and $\lambda_2=788.28$~nm, atoms in state $\left\vert 1 \right\rangle$ and $\left\vert 2 \right\rangle$, are diffracted by the associated optical lattice $V_1$ and $V_2$, respectively.}
\label{Fig2}
%}
\end{figure}

The two spin states of $^{87}$Rb atoms constitute the synthetic dimension that accommodates the two twisted layers of lattices $V_1$ and $V_2$. To precisely determine the tune-out wavelengths $\lambda_1$ and $\lambda_2$ of the optical lattices $V_{1}$ and $V_{2}$, we measure the diffraction of atoms by the optical lattices. The experimental sequence starts with an almost pure BEC in a crossed-beam dipole trap. The atoms are prepared in one of the two spin states and a short pulse of the lattice beams is applied. The lattice potential induces Bragg diffraction of atoms to high momentum states. After turning off the lattice beams, we image the diffracted atoms. The wavelengths of the lattice beams are finely adjusted to the ``tune-out" wavelengths such that atoms in state $\left\vert 1 \right\rangle$ are only diffracted by the lattice potential $V_1$ and not by the potential $V_2$ as shown in Fig. 2. Similarly atoms in state $\left\vert 2 \right\rangle$ only experience the potential $V_2$, but not $V_1$. By eliminating the cross-talks, we determined the ``tune-out" wavelengths to be $\lambda_1=790.02$~nm and $\lambda_2=788.28$~nm. Remarkably the lattice beams are circularly polarized to produce spatial intensity modulation such that the lattice potentials are attractive to atoms in both spin states (see Supplementary Material section II for details).

\begin{figure}
\includegraphics[width=3.5in]{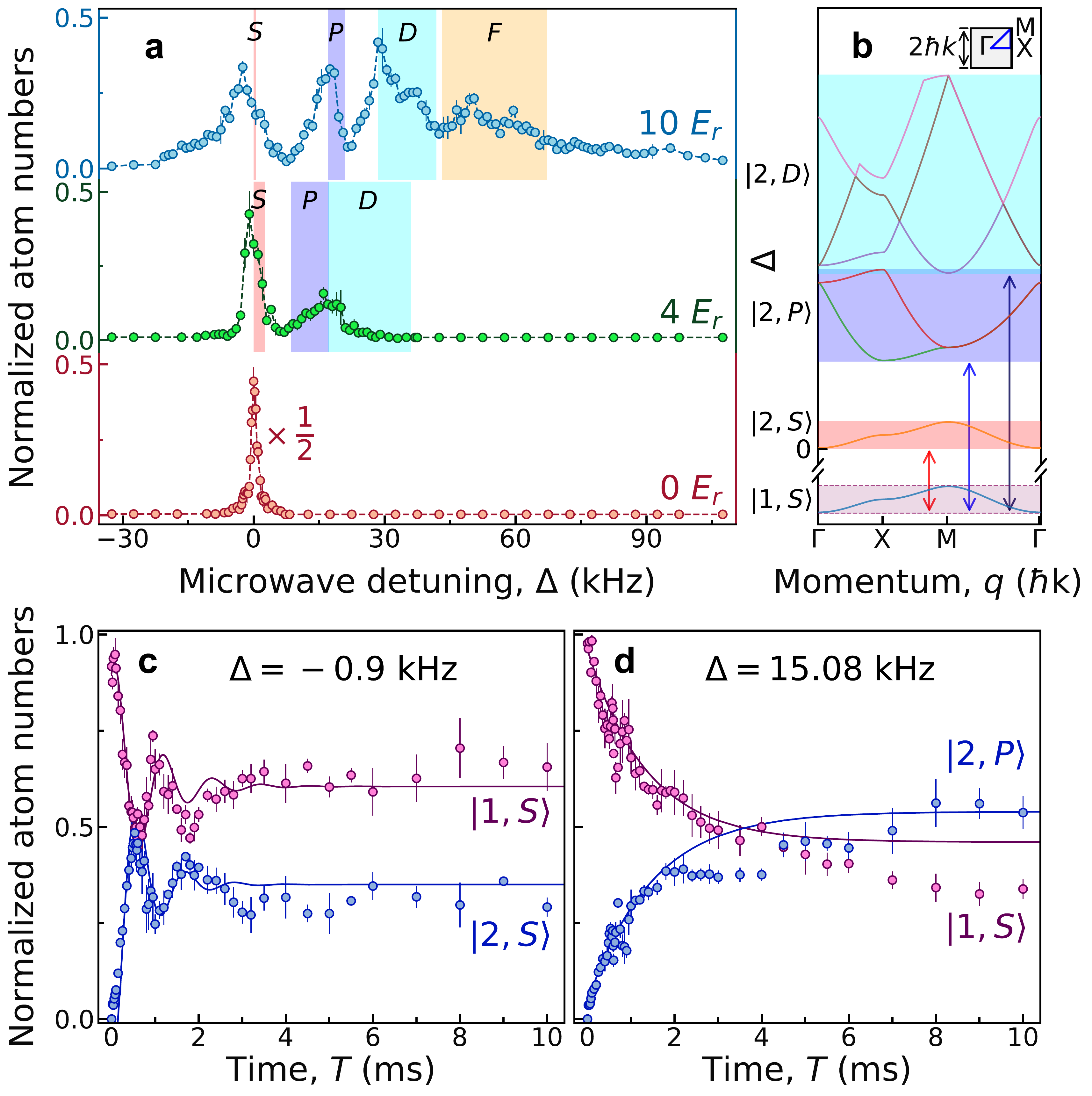}
%\begin{figure*}
%\centerline{
%\includegraphics[width=0.45\linewidth]{fig1.pdf}
%} \vspace{0.1in}
%\setlength{\columnwidth}{3.2in}
%\centerline{
\caption{ \textbf{Interlayer coupling in twisted-bilayer optical lattices.}
$\bf{a}$~MW spectrum of atoms in the twisted-bilayer optical lattices. Atoms in spin state $\left\vert 1 \right\rangle$ are driven by MW to spin state $\left\vert 2 \right\rangle$ in the presence of the lattice potential with depth of 0 (no lattice), 4 and 10$E_r$. Here $E_{r}=q_{r}^2/2m = h\times3.67$~kHz is the recoil energy, $q_{r}=\hbar k=h /\lambda$ is the recoil momentum, $m$ is the atomic mass of $^{87}$Rb, and $\lambda$ is the wavelength of the lattice laser. The MW pulse length of 530~$\mu$s corresponds to a $\pi$ pulse in the absence of the lattice potential. $\bf{b}$~Lattice band structure for the two spin states calculated with the lattice depth 4$E_r$. The MW field drives atoms from the $s$-band of state $\left\vert 1 \right\rangle$, labelled as $|1,S\rangle$ to $s$-, $p$- and $d$-bands of state $\left\vert 2 \right\rangle$ with different detuning $\Delta$. $\bf{c}$ and $\bf{d}$~Starting with all atoms in $|1,S\rangle$, population in state $\left\vert 2 \right\rangle$ is measured in the twisted-bilayer lattices at 4$E_r$ after the MW pulse that drives the atoms to $|2,S\rangle$ with the detuning $\Delta=-0.9$~kHz, or to $|2,P\rangle$ with detuning $\Delta=15.08$~kHz. Fits in panel $\bf{c}$ show an interlayer coupling frequency of $\Omega_\mathrm{R}=$2$\pi\times 893$~Hz and a decay rate of 1200/s. Lines in panel $\bf{d}$ are guides to the eye. Each point is based on three or more measurements and error bars show the standard deviations of the mean.}
\label{Fig3}
%}
%\end{figure*}
\end{figure}

Experimentally intralayer hoppings $t_1$ and $t_2$ between lattice sites are controlled by the depth of the optical lattices $V_1$ and $V_2$; interlayer hopping $\Omega_{\mathrm{R}}$, on the other hand, is independently induced by microwaves (MW) that couple the two spin states. Starting with atoms in state $\left\vert 1 \right\rangle$ in the dipole trap, for example, the MW spectrum displays a single narrow peak when atoms are driven to state $\left\vert 2 \right\rangle$. By loading the atoms into the twisted-bilayer optical lattices, the spectrum displays multiple peaks. The peaks correspond to transitions from atoms in the ground band of lattice $V_1$, which we label $|1,S\rangle$, to different Bloch bands of lattice $V_2$, which we label $|2, S\rangle$, $|2, P\rangle$, $|2, D\rangle$ and so on, see Figs.~3a and 3b. The peak locations agree with the calculated energies of the $s$-, $p$- and $d$-bands in lattice $V_2$. Remarkably, the multi-peak structure supports that atoms in different spin states are confined in different lattices. If atoms are loaded into a spin-independent lattice, only a single narrow peak shows up in the spectrum, which belongs to the $|1,S\rangle$ to $|2,S\rangle$ transition. This is because MW transitions between different Bloch bands are negligible in spin-independent lattices. In the twisted optical lattice, the transitions from $s$-band of state $\left\vert 1 \right\rangle$ to other bands of state $\left\vert 2 \right\rangle$ are allowed. In the presence of the twisted bilayer lattices, the transitions are broadened since the two spin states experience different trap potentials, which induce fast dephasing. Moreover, the on-site interactions increase in deeper lattice potential, resulting in faster decay from high bands to lower bands and thus broader spectral lines. Our observation supports MW as a versatile and powerful tool to induce interlayer hopping between the two twisted layers in the synthetic (spin) dimension.

To quantify the interlayer hopping energy, we measure the time evolution of the population in state $\left\vert 2 \right\rangle$. We observe a coherent oscillation at detuning $\Delta=-0.9$ kHz, which corresponds to the transition from $|1,S\rangle$ to $|2, S\rangle$, see Fig.~3c. The interlayer coupling strength can be determined from the oscillation frequencies. In our experiment, the coupling strength is tunable up to 1$E_{r}$, which exceeds that in typical twisted-bilayer graphene systems. On the other hand, coupling to the $p$-band $|2,P\rangle$ leads to faster decay likely due to collisional relaxation to the lower $s$-band, see Fig.~3d. In the following, we will focus on atoms in the twisted-bilayer optical lattices with MW-induced coupling between the $s$-bands of the two layers.

A key signature of atoms in the twisted-bilayer optical lattice is the moir$\acute{\mathrm{e}}$ lattice with a period

\begin{equation}\label{eq:1}
	\begin{aligned}
	\lambda_{mo}=\frac a{2\sin\theta/2},\\
	\end{aligned}
\end{equation}
which, for the lattice constant $a=395$~nm and twist angle $\theta=5.21^{\circ}$, amounts to $\lambda_{mo}=4.35$~$\mu$m. The large moir$\acute{\mathrm{e}}$ period gives rise to a mini-Brillouin zone in the momentum space, which is expected to generate the flat bands and strongly correlated states~\cite{Cao2018-1,Cao2018-2,Yankowitz1059,Lu2019,WANG2019100099,Andrei2020,Balents2020,Kennes2021,PhysRevLett.99.256802,PhysRevB.81.161405,Bistritzer12233,PhysRevB.85.195458}. Notably \emph{in-situ} moir$\acute{\mathrm{e}}$ pattern is also observed in 1D lattices with two lattice constants~\cite{PhysRevX.9.021001}. To identify the moir$\acute{\mathrm{e}}$ length scale in our system, we employ \emph{in situ} absorption imaging to visualize the moir$\acute{\mathrm{e}}$ pattern, see Fig.~4. Here we first load the atoms in state $\left\vert 1 \right\rangle$ into the lowest $s$-band of lattice $V_1$ and then ramp up the MW field with detuning $\Delta=-0.9$ kHz to drive the transition from $|1,S\rangle$ to $|2,S\rangle$. We then \emph{in situ} image the atoms in state $\left\vert 2 \right\rangle$. Moir$\acute{\mathrm{e}}$ patterns in one and two dimensions are observed, and the moir$\acute{\mathrm{e}}$ period is measured to be 4.35~$\mu$m consistent with expectation, see Figs.~4a-f. Note that the primary optical lattice spacing $a=395$~nm is indiscernible with our imaging optics.

We also examine the quantum state of atoms in the bilayer twisted lattices by analyzing their momentum space distribution. After loading a BEC into the bilayer lattice of 4$E_r$ in the presence of resonant MW transition, we hold for some time and then perform the time-of-flight measurement, see Figs.~4g and 4h. Two sets of diffractions manifest, which correspond to the primary lattice momentum $\pi/a$ and the much smaller moir$\acute{\mathrm{e}}$ momentum $\pi/\lambda_{mo}$. The high contrast of both sets of diffraction pattern suggests that the atoms remain in the superfluid phase with phase coherence extending beyond the moir$\acute{\mathrm{e}}$ length scale. In particular, the contrasts of the moir$\acute{\mathrm{e}}$ pattern in real and momentum space persist over 40~ms, see Fig.~4i, from which we conclude that the atoms maintain in the superfluid phase in the twisted-bilayer lattices.

\begin{figure*}
\includegraphics[width=7in]{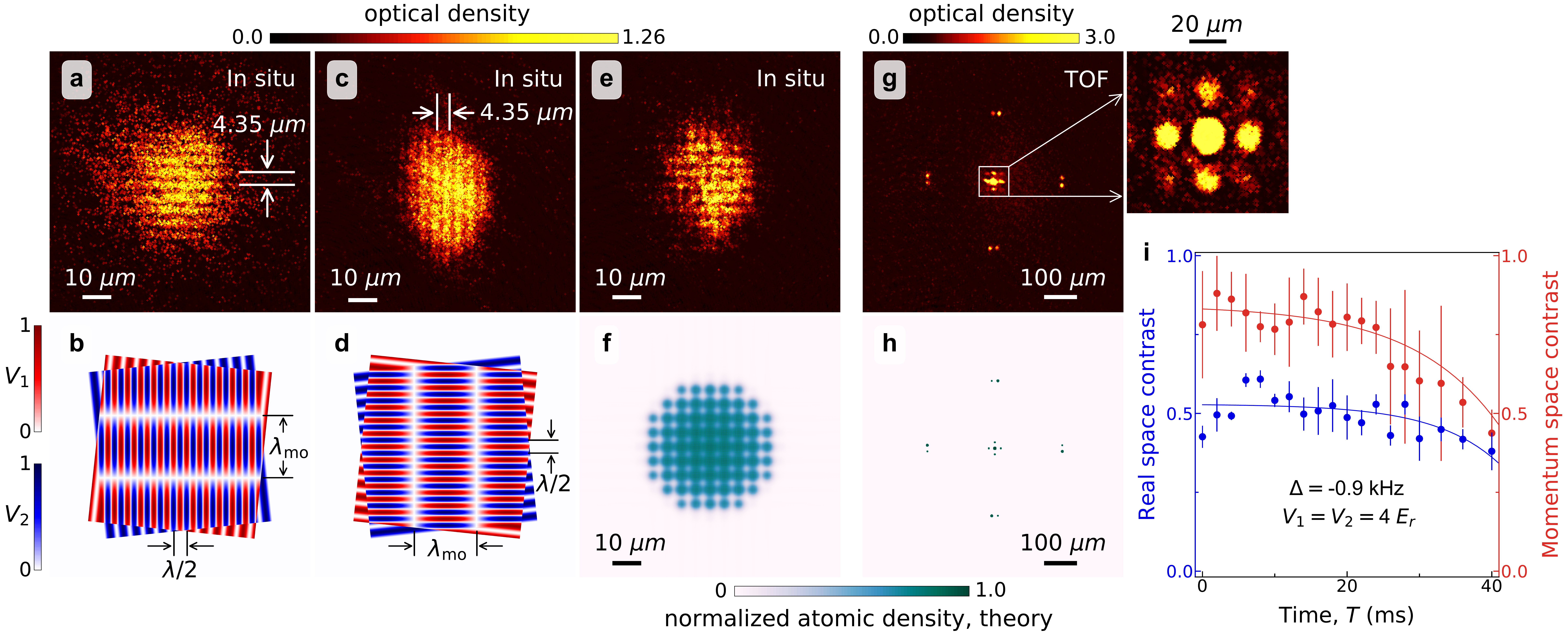}
%\begin{figure*}
%\centerline{
%\includegraphics[width=0.45\linewidth]{fig1.pdf}
%} \vspace{0.1in}
%\setlength{\columnwidth}{3.2in}
%\centerline{
\caption{ \textbf{Moir$\acute{\mathrm{e}}$ pattern and superfluid ground state in twisted-bilayer optical lattices.}
$\bf{a}$ and $\bf{c}$ Moir$\acute{\mathrm{e}}$ pattern of atoms in one dimensional bilayer optical lattices in the $x$- and $y$- direction. $\bf{b}$ and $\bf{d}$ Plot of the corresponding optical lattice potential. Note that the primary optical lattice $\lambda/2$ is indiscernible in the \emph{in situ} images. In all experiments, an MW field is applied to couple the $s$-band of state $\left\vert 1 \right\rangle$ and $s$-band of state $\left\vert 2 \right\rangle$. The atoms in state $\left\vert 2 \right\rangle$ are measured. The elliptical shape of the atomic cloud is induced by a little asymmetric harmonic trap potential in x-y plane. $\bf{e}$ and $\bf{f}$~Experimental and theoretical moir$\acute{\mathrm{e}}$ pattern of atoms in the presence of optical lattices in both directions. $\bf{g}$ and $\bf{h}$ Time-of-flight (TOF) images with 18~ms from the experiment and calculation show diffraction peaks associated with the primary lattice constant $\lambda/2$ and moir$\acute{\mathrm{e}}$ length $\lambda_{mo}$. The enlarged picture of the center part of $\bf{g}$ is obtained by the imaging system with higher magnification. $\bf{i}$ The contrast of moir$\acute{\mathrm{e}}$ pattern in real space and the contrast of diffraction pattern for different hold times. Here, the lattice depths are $V_{1x}=V_{2x}=V_{1y}=V_{2y}=4E_r$ ($U/t=1.67$, $\Omega_\mathrm{R}/t=2.07$, $U$ is the on-site interaction). The real-space and momentum-space distributions of $\bf{f}$ and $\bf{h}$ are theoretically calculated by solving the mean-field ground states according to the Gross-Pitaevskii equations (see Supplementary Material section IV for details). The contrast in the real space is defined as $(S_{max}-S_{min})/(S_{max}+S_{min})$, where $S_{max}$ and $S_{min}$ are the maximum and minimum atomic density of the moir$\acute{\mathrm{e}}$ fringes. The contrast in the momentum space is defined as $(P^{av}_{max}-P_{min})/(P^{av}_{max}+P_{min})$, where, $P^{av}_{max}=(P^{0}_{max}+P^{1}_{max})/2$ and $P_{min}$ are the average maximum and minimum density of the diffraction pattern. Here $P^{0}_{max}$ is zero momentum component and $P^{1}_{max}$ is the moir$\acute{\mathrm{e}}$ component near the zero momentum. Error bars show the standard deviation of the mean. }
\label{Fig4}
%}
%\end{figure*}
\end{figure*}

Theoretically, depending on the twist angle $\theta$, the superimposed twisted-bilayer lattice can yield either a periodic potential with supercells that supports a delocalized ground state or a quasi-periodic one that supports a localized ground state in the absence of interactions. In fact, only specific twist angles give rise to periodic lattice potentials. For square lattices, the twist angles that lead to commensurate superlattice should satisfy $\theta=2\arctan(\bar{m}/\bar{n})$, where $\bar{m}$ and $\bar{n}$ are integers~\cite{Wang2020}. The twist angle $\theta=5.21^{\circ}$ used in our work is close to the commensurate angle $\theta=2\arctan(1/22)\approx 5.205^{\circ}$, and the period of the supercell is given by $22a=2\lambda_{mo}$ (see Supplementary Material section III for details). While our twist angle does not exactly match the commensurate angle $\theta=2\arctan(1/22)\approx 5.205^{\circ}$, the small difference can not be distinguished in a finite size sample due to repulsive interactions (see Supplementary Material section IV for details). In particular, the spatial moir$\acute{\mathrm{e}}$ period remains a clear observable in our experiment because of the finite chemical potential of our atomic superfluid. The persistence of the spatial and momentum periodicity of the sample in the twisted-bilayer lattice supports the superfluid as the ground state of the system.

Compared with electronic materials, where the flat band is investigated frequently near the Fermi surface, we can also explore flatband physics with bosons condensed in the lowest band. In our system, when interlayer coupling increases, the long-wavelength moir$\acute{\mathrm{e}}$ potential becomes deeper, so atoms in the lowest band are isolated at a larger spatial scale (moir$\acute{\mathrm{e}}$ wavelength), which
flattens the ground band and enhances the localization of the atoms. In the large interlayer coupling limit, the system can be regarded as a single-layer (single-component) experiencing a twisted optical lattice
(see Supplementary Material section III for details). The single-layer system with a twisted optical lattice admits
a flatband structure in the ground band, which has also been studied experimentally in photonic systems~\cite{Wang2020,Huang2016,Fu2020}. The easily-tuned intra- and interlayer couplings in our system offer the added advantages of seeking novel quantum phases and phase transitions with cold atoms.

By varying the depth of optical lattices and interlayer coupling, we find several distinct quantum phases, including superfluid (SF), superfluid with only short-range coherence (SF-II), Mott Insulator (MI) and insulator (I), (see Fig.~5a and Supplementary Material section IV). These phases can be distinguished by the phase coherence and real-space density correlations. Remarkably the SF-II phase emerges with finite interlayer coupling around the transition from a regular SF to an insulator. The spatial range of phase coherence is the key to distinguishing the two SF phases: while an SF supports long-range phase coherence \cite{PhysRevA.72.053606}, the SF-II phase maintains the coherence only up to the moir$\acute{\mathrm{e}}$ length scale. In addition, the SF-II phase supports the moir$\acute{\mathrm{e}}$ pattern in the real-space. Theoretically, SF-II is the phase with superfluid domains embedded in a gapped insulator, induced by the interlayer coupling. Finally, the insulator phases I and MI can be identified by the disappearance of spatial coherence at all scales and integer fillings of all the sites. While the MI has uniform atom density with weak interlayer coupling, the I phase features moir$\acute{\mathrm{e}}$ pattern due to stronger interlayer coupling. % So SF-II phase can be identified by checking the disappearance of the moir$\acute{\mathrm{e}}$-scale long-range correlation with vanishing moir$\acute{\mathrm{e}}$ lattice momentum but the remaining of the shore-range coherence with residual primary lattice momentum. %

\begin{figure*}
\includegraphics[width=7in]{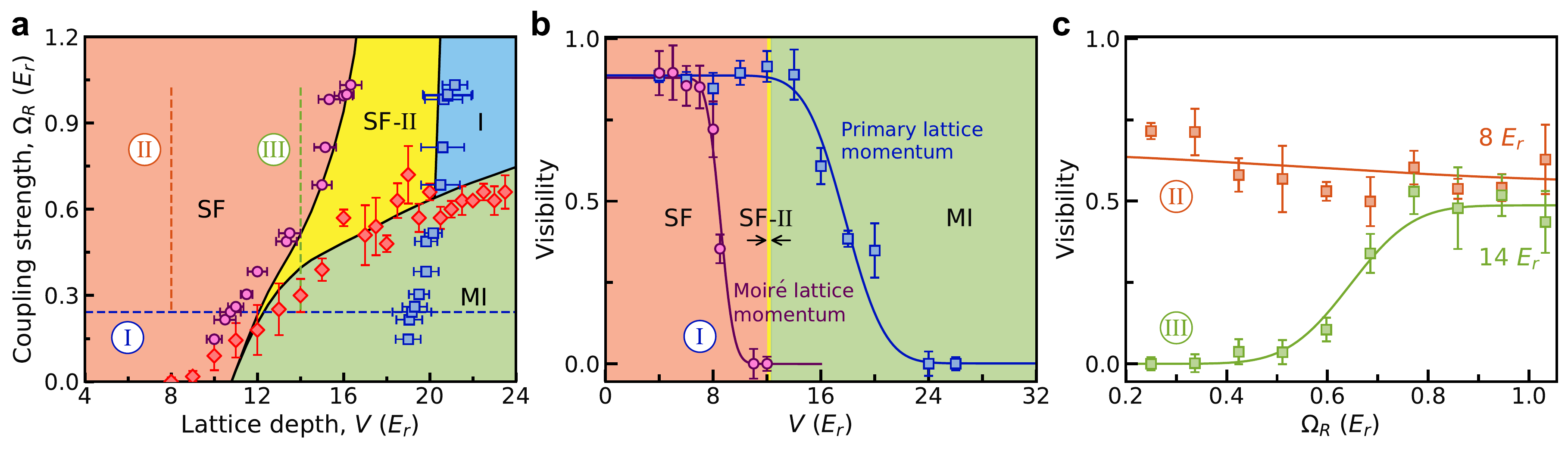}
%\begin{figure}
%\centerline{
%\includegraphics[width=0.45\linewidth]{fig1.pdf}
%} \vspace{0.1in}
%\setlength{\columnwidth}{3.2in}
%\centerline{
\caption{ \textbf{Phase transition for the twisted-bilayer optical lattice.}
$\bf{a}$ Phase diagram (see Supplementary Material section IV for details), where SF, SF-II, MI, and I refer to superfluid, superfluid only with short-range coherence, Mott insulator, and insulator. The solid curves denote the calculated phase boundaries with mean-field at zero temperature. The dots are experimental measurements of the phase boundaries. The pink circles and blue squares denote the loss of the coherence at the moir$\acute{\mathrm{e}}$ length scale and the length scale of the primary lattice respectively. The red diamonds denote the appearance of the moir$\acute{\mathrm{e}}$ pattern in the real-space (in situ images) as increases the interlayer coupling. The error bars indicate the experimental uncertainties in determining the phase transition. $\bf{b}$ Visibility curves for the moir$\acute{\mathrm{e}}$ and primary lattice momentum components as a function of lattice depth (Path I in $\bf{a}$). A sequential loss of phase coherence appears at the moir$\acute{\mathrm{e}}$ momentum and the primary lattice momentum. The intermediate regime indicates SF-II phase. The interlayer coupling frequency is $\Omega_\mathrm{R}=0.24E_{r}$. $\bf{c}$ Visibility curves for the moir$\acute{\mathrm{e}}$ momentum component as a function of interlayer coupling with the lattice depth $8E_{r}$ (Path II in $\bf{a}$) and $14E_{r}$ (Path III in $\bf{a}$) respectively. The solid lines in $\bf{b}$ and $\bf{c}$ are fitted from the experimental data and only guide the eye.}
\label{Fig5}
%}
\end{figure*}
%\end{figure}

In the experiment, we measure the phase coherence from the momentum-space diffraction peaks in the TOF images and directly probe the  moir$\acute{\mathrm{e}}$ pattern by \emph{in situ} imaging following the measurement method as shown in Fig. 4. The measurement of the phase boundaries is shown in Fig. 5a. We employ three independent paths to study these phases. Path I, we fix the interlayer coupling strength at a small value $\Omega_R =0.24E_{r}$ and increase the lattice depth. The phase transition from SF to MI and across SF-II is shown in the time-of-flight images, see Fig. 5b. Here the diffraction peaks at the moir$\acute{\mathrm{e}}$ momenta disappear first before the disappearance of the primary lattice. The intermediate regime indicates the SF-II phase, where the moir$\acute{\mathrm{e}}$-scale long-range correlation is destroyed while a short-range coherence remains, at the same time, the density correlations of moir$\acute{\mathrm{e}}$ pattern appears in the real-space. Path II, we fix lattice depth in the SF region and increase the interlayer coupling. The diffraction peaks at the moir$\acute{\mathrm{e}}$ momenta persist with high contrasts. However, in Path III, when the depth of optical lattices is fixed at the MI region and the interlayer coupling increases, the visibility at the moir$\acute{\mathrm{e}}$ momenta presents the threshold behavior and emerges at $\Omega_R >0.5E_{r}$, see Fig.~5c. These observations are qualitatively consistent with the theoretical expectation and demonstrate that the interlayer coupling can induce a reentrant transition from MI to SF across SF-II. One may understand such rich transitions from the fact that the interplay between the interlayer coupling and interactions tends to localize the bosons, primarily in the moir$\acute{\mathrm{e}}$ length scale.

This work provides a preliminary physical insight into the quantum phase transition between SF and SF-II ( MI and SF-II or MI and I) and offers the possibility to study the complex phases due to the presence of quasi-disorder induced by large interlayer coupling and strong interaction, such as Bose glass insulator, resembling that in disordered bosonic systems \cite{PhysRevB.40.546,PhysRevLett.67.2307,PhysRevB.53.2691}. These complex phases are worth further investigating in the future.

% An equivalent condition is $cos\theta=a/c$ and $sin\theta=b/c$, which can be defined by Pythagorean triples (that is $a^{2}+b^{2}=c^{2}$, $(a,b,c)\in N$ are positive integers). The relationship between (m,n) and (a,b,c) is $(m+in)^{2}=(a+ib)$,($a=m^{2}+n^{2}$, $b=2mn$). The first form involving half angles is in better illumination with the spinorial character of the rotation.

% We can directly observe the spatial moir$\acute{\mathrm{e}}$ pattern and the momentum wavefunction, which confirms superfluidity in the bilayer lattices.

%Since the transition from $|1,S\rangle$ to $|2, P\rangle$ (or higher orbital bands) is easily implemented in this work, optical lattice simulator beyond conventional $s$-band Hubbard physics can be realized with orbital degrees of freedom by making use of higher Bloch bands \cite{ Li_2016}. The higher orbital bands in the twisted-bilayer lattices can be used to construct quantum simulators of exotic models beyond natural crystals, complex BECs and topological materials.

The present work focuses on the realization and the ground state properties of atoms in the twisted-bilayer optical square lattice. Our success in loading a superfluid into the bilayer lattice demonstrates a new versatile platform to explore moir$\acute{\mathrm{e}}$ physics and the associated superfluidity in a quantum many-body system. Beyond the tunable twist angle, the cold atom platform offers remarkable controls such as different lattice depths and interlayer coupling in different layers.

Furthermore, the twisted-bilayer square lattice closely connects to the physics of heterostructures of twisted atomically thin semiconductors~\cite{PhysRevResearch.1.033076, Kennes2020,Kennes2021}. At the same time, our experiment can in principle be extended to multi-layer lattice where the interlayer couplings can be independently induced by MW and radio-frequencies. Replacing the MW by optical Raman transitions, the interlayer coupling can be spatial dependence, which can support topological ground states. Finally, our optical lattice scheme can be applied to confine fermionic atoms in bilayer hexagonal lattice, which faithfully simulates electrons in a bilayer graphene, and may offer insight into the emergence of superconductivity in the strongly correlated, flatband regime.

\section*{DATA AVAILABILITY}
All data generated or analysed during this study are included in this published article. Additional data are also available from the corresponding authors upon reasonable request.

\begin{acknowledgments}
This research is supported by Innovation Program for Quantum Science and Technology (Grant No. 2021ZD0302003), National Key Research and Development Program of China (2016YFA0301602, 2018YFA0307601, 2022YFA1404101), NSFC (Grant No. 12034011, 12022406, 11804203),
the Fund for
Shanxi ``1331 Project" Key Subjects Construction and Tencent (Xplorer Prize). CC acknowledges support by the National Science Foundation (Grant
No. PHY-2103542), and the Army Research Office
STIR (Grant W911NF2110108).

\end{acknowledgments}

\section*{Author contributions}
J.Z. conceived the idea and performed the experimental designs. L.W., Z.M., F.L., K.W., P.W. and J.Z. performed the experiments. C.C., Z.M., L.W., F.L., W.H. and J.Z. analysed the data and all authors discussed the results. W.H., C.G. and J.Z. performed the simulation.  Z.M. plotted the figures. J.Z. and C.C. wrote the manuscript. All authors interpreted the results and reviewed the manuscript. J.Z. designed and supervised the project.

\section*{Competing interests}
The authors declare no competing financial interests.

\bibliography{references}
\bibliographystyle{naturemag}

\subsection*{Supplementary Material}

\noindent \textbf{I. Experimental setup}

In our experiment, the ultracold $^{87}$Rb atoms in the $|F=2,m_{F}=2\rangle$ state are prepared in the crossed optical dipole trap \cite{Xiong:10}. Forced evaporation in the optical trap creates the BEC with up to $5\times10^5$ atoms. The atoms
can be transferred to the $|F=1,m_{F}=1\rangle$ state via a rapid adiabatic passage induced by microwave transition. To load the atoms into the 2D trap, a 532~nm laser beam is deflected by an acousto-optic deflector (AOD) and then split into two beams with variable spacing adjusted by the AOD. The two beams are focused onto the atoms with a 150~mm aspherical lens. These beams interfere to form a standing wave in the vertical direction with variable separation (Accordion lattice). This separation can be varied from 12~$\mu m$ down to 3~$\mu m$. The advantage of variable spacing is that we can load a 3D shaped cloud into a single layer of the 2D pancakes at maximum separation and then compress the pancake adiabatically to reach a deep 2D regime. The maximum vertical confinement can reach more than 20~kHz and we optimize it at 1~kHz to observe moir$\acute{\mathrm{e}}$ pattern and superfluid of ultracold atoms.

The twisted-bilayer optical lattices are created by two sets of 2D square lattice $V_{1}$ and $V_{2}$. A twisted angle of $\theta=5.21^{\circ}$ is set between the two lattice potentials, namely, $V_{2}(\bf{r})$ $=V_{1}(S\bf{r})$, $S= \left(
      \begin{array}{cc}
\cos\theta & -\sin\theta \\
        \sin\theta &\cos\theta \\
       \end{array}
    \right)$. The optical lattices $V_{1}$ and $V_{2}$ are derived from two CW Ti:Sapphire single
frequency lasers (M Squared lasers SolsTiS and Coherent MBR-110) respectively. Two lattice beams $V_{1x}$ and $V_{1y}$ of $V_{1}$ are frequency-shifted +80~MHz and +95~MHz by two single-pass acousto-optic
modulators (AOMs), respectively. The same applies to the two lattice beams $V_{2x}$ and $V_{2y}$ of lattice $V_{2}$. The four lattice beams are coupled into polarization-maintaining single-mode fibers in order to improve the stability of the beam pointing and achieve better
beam-profile quality. After the fibers, each lattice beam is focused by a lens and retroreflected by a concave mirror. In order to generate the vector light shift, we use the same circular polarization for two lattice beams to produce spatial intensity modulation. In the experiment, we can determine and calibrate this angle by measuring the intersection angle between two lasers and the moir$\acute{\mathrm{e}}$ period from the \emph{in situ} images. The estimated uncertainty of the two methods is about 0.05$^{\circ}$.

We use the MW field to couple the two spin states for manipulating the interlayer coupling. The 6.8 GHz MW signal is amplified by a 10 W solid state amplifier (Kuhne Electronic, KU PA 640720-10A). We place a circulator on the output of the amplifier to reduce reflected power coming back to the amplifier. The MW is emitted out to the atoms by a sawed-off waveguide, which is placed outside of the high vacuum glass cell. We use MW cables to transfer MW from the amplifier to the waveguide. With this MW power amplifier, we can reach the maximum interlayer coupling strength of about 1.0$E_{r}$. It is feasible to increase the interlayer coupling strength to about several $E_{r}$ by using an available higher power amplifier.

Our image system consists of an objective with a numerical aperture of NA=0.69, working distance of 11~mm and effective focal length of 18~mm. A 900~mm lens after the objective leads to a magnification of 50 for \emph{in situ} imaging with an EMCCD (Andor iXon Ultra 897). We also employ a 200~mm (400~mm) lens after the objective leads to a magnification of 11 (22) for the time-of-flight absorption imaging with 18 ms. The atoms are detected by state-selective absorptive imaging. Since we choose two ground hyperfine Zeeman states of $^{87}$Rb $\left\vert F=2,m_{F}=0 \right\rangle$ of the $F=2$, and $\left\vert 1,1\right\rangle$ of the $F=1$ hyperfine manifold as the two internal spin states, we can fully resolve the population in each individual state. For $\left\vert F=2, m_F=0\right\rangle$ state, a 50 $\mu$s long imaging pulse of resonant light on the $F=2\rightarrow F'=3$ D$_{2}$ cycling transition is used to detect the $\left\vert 2\right\rangle$ atoms. In order to detect $\left\vert F=1, m_F=1\right\rangle$ state, a resonant light pulse on the $F=2\rightarrow F'=3$ cycling transition is firstly used to remove the $\left\vert 2\right\rangle$ atoms and then a 50 $\mu$s long imaging pulse of resonant light on the $F=2\rightarrow F'=3$ is applied at the same time with a repump light (resonant light $F=1\rightarrow F'=2$) to detect the $\left\vert 1\right\rangle$ atoms.

When studying the superfluid-to-Mott insulator transition, we use the standard method of interference pattern contrast (visibility) to reveal this
transition \cite{PhysRevA.72.053606,Greiner2002}. We first load the atoms in state $\left\vert 1 \right\rangle$ into the lowest $s$-band of lattice $V_1$ by ramping up $V_1$ and $V_2$ simultaneously with 30 ms, and then ramp up the MW field with 10 ms to drive the transition from $|1,S\rangle$ to $|2,S\rangle$. The atoms are detected by state-selective absorptive imaging with time-of-flight of 18 ms after switching off all lattices and trapping light. In experiment, we first check that BEC is still kept (see Fig. S1) as ramp up the lattice $V_1$ (or $V_2$) to the higher lattice depth than $24E_{r}$ and then ramp down again, which makes sure to perform the phase transition from SF to MI successfully for the lattice V1 (or V2). When adding the interlayer coupling between two spin states, and at the same time a quasi-disorder is introduced, there are two more mechanics to make the system not completely reversible. One is the finite coherent time between two spin states. When the system is prepared initially in the spin down, the system will become the spin mixture after the interlayer coupling is ramped back down. We define this process as irreversibility. The other is that adiabaticity is broken down by a quasi-period or disordered lattice, which induces not to completely remain in the zero-momentum state after ramping the lattices back down.

\begin{figure}[tb]
\includegraphics[width=0.48\textwidth]{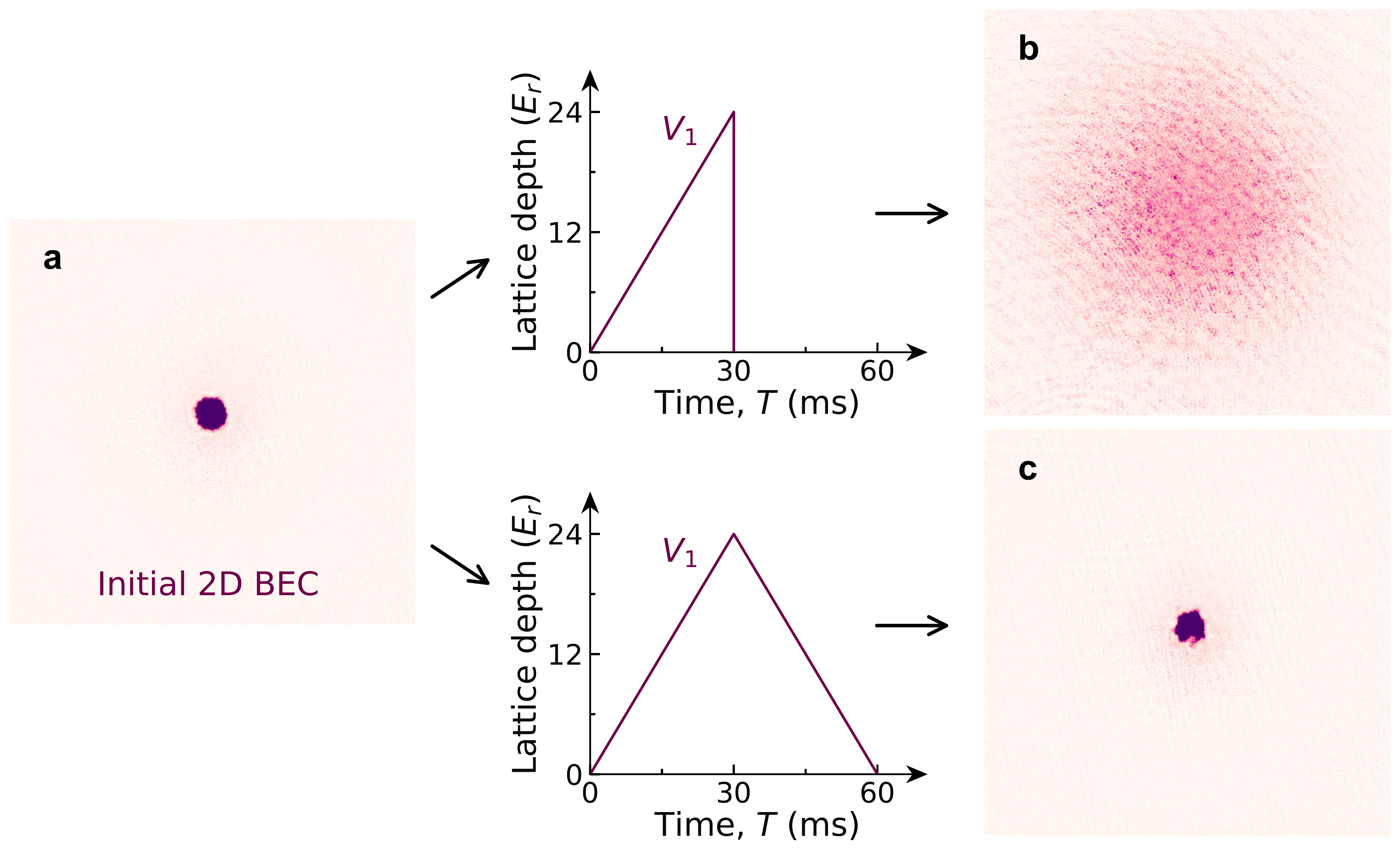}
\setcounter{figure}{0}
\renewcommand{\thefigure}{S\arabic{figure}}
\caption{ \textbf{Coherence in the SF-MI transition.}
$\bf{a}$ The initial BEC in 2D pancake-like potential. $\bf{b}$ Absorption image after atoms are released abruptly from an optical lattice potential $V_1$ (or $V_2$) with a potential depth $24E_{r}$. $\bf{c}$ Absorption image when the lattice is ramped up to the lattice depth $24E_{r}$ and then ramp down to zero. The images are obtained after 18 ms free space expansion.}
\label{sFig1}
\end{figure}

\noindent \textbf{II. Tune-out wavelength for twisted-bilayer optical lattices}

\begin{figure*}[tb]
\includegraphics[width=6.9in]{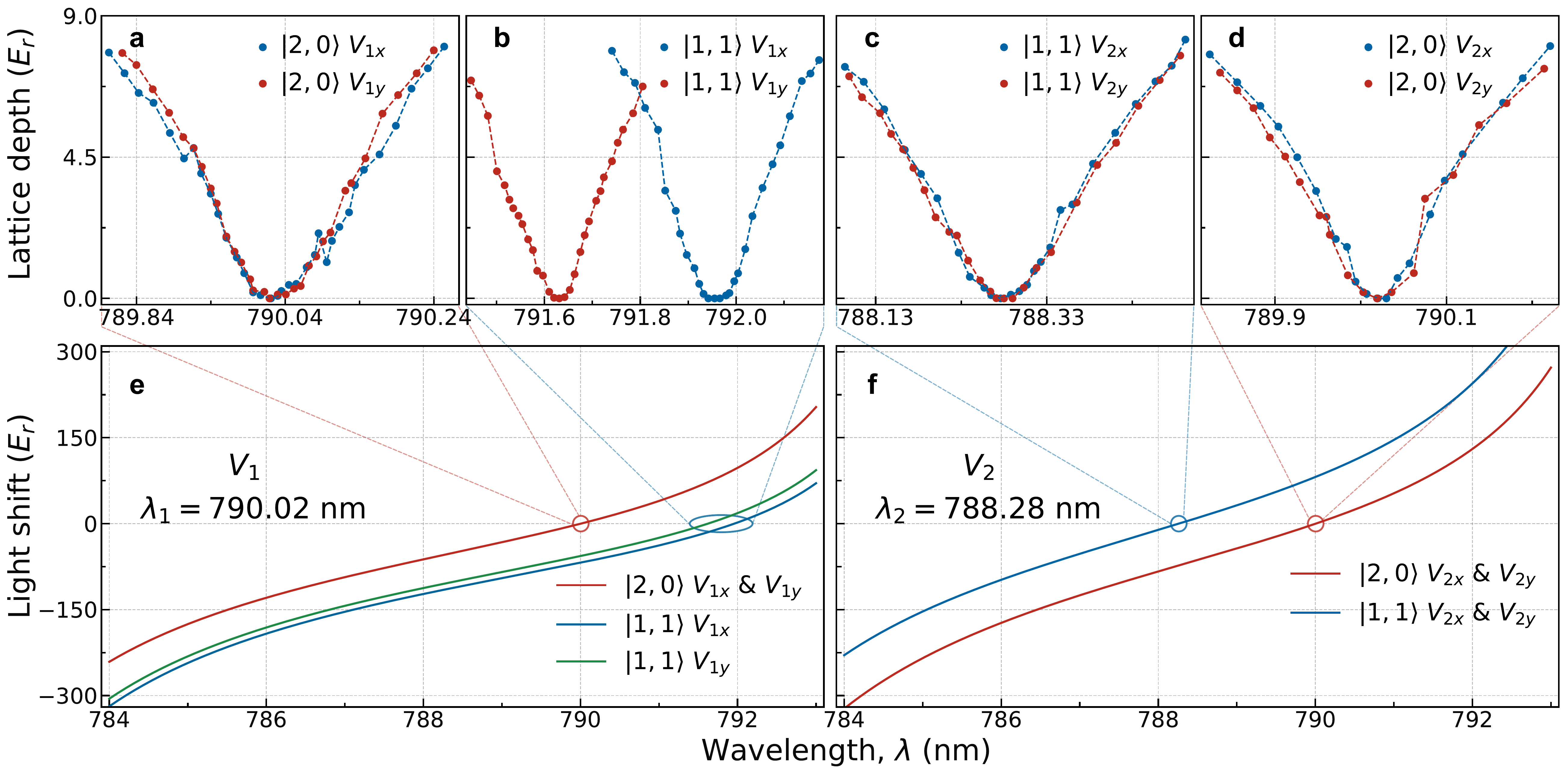}
\setcounter{figure}{1}
\renewcommand{\thefigure}{S\arabic{figure}}
\caption{ \textbf{Determination of tune-out wavelengths.}
$\bf{a}$-$\bf{b}$ The lattice depth $V_{1x}$ (blue) and $V_{1y}$ (red) as a function of wavelength $\lambda$ for the two different hyperfine states $|F=1,{{m}_{F}}=1\rangle$ and $|F=2,{{m}_{F}}=0\rangle$. The angles between $V_{1x}$, $V_{1y}$ and $B_{0}$ are $39.79^{\circ}$ and $50.21^{\circ}$ respectively. $\bf{c}$-$\bf{d}$ The potential depth $V_{2x}$ (blue) and $V_{2y}$ (red) as a function of wavelength $\lambda$ for the two different hyperfine states $|F=1,{{m}_{F}}=1\rangle$ and $|F=2,{{m}_{F}}=0\rangle$. $\bf{e}$~Theoretical light shift of $V_{1x}$, $V_{1y}$ for $|1, 1\rangle$ and $|2, 0\rangle$. $\bf{f}$~Theoretical lattice depth of $V_{2x}$, $V_{2y}$ for $|1, 1\rangle$ and $|2, 0\rangle$. The bias magnetic field of 10~Gauss is applied along the $45^{\circ}$ diagonal line of the square lattice $V_{2}$.}
\label{sFig1}
\end{figure*}

AC Stark shift, or light shift is a light-induced change of energy level. For alkali-metal atoms, the total ac Stark shift can be expressed in the irreducible components (including scalar, vector, tensor components) of the polarizability
\cite{steck2007quantum}:
\begin{equation}\label{eq:1}
	\begin{aligned}
		\Delta U=&\Delta U\left( F,{{m}_{F}};\omega  \right)\\
		=&-I\Big[\alpha _{~}^{\left( 0 \right)}\left(\omega  \right)
		+{{\alpha }^{\left( 1 \right)}}\left(\omega  \right) \left( \xi \hat{e}_{k}\cdot\hat{e}_{B} \right) \frac{{m}_{F}}{F}\\
		& + {{\alpha }^{\left( 2 \right)}}\left(\omega  \right) \frac{3\cos^{2}\phi-1}{2} \frac{3{m}_{F}^{2}-F(F+1)}{F(2F-1)}\Big],\\
	\end{aligned}
\end{equation}
where $F$ is the total atomic angular momentum, $m_{F}$ is the magnetic quantum number, $\omega$ is the laser frequency, $I$ is the laser field intensity, $\xi$ is light ellipticity, $\hat{e}_{k}$ and $\hat{e}_{B}$ are unit vectors along the light wave-vector and magnetic field quantization axis respectively, $\phi$ is the intersection angle between the linearly polarized component of light field and $\hat{e}_{B}$. This formula comes from the perturbation expansion. Note that the range of values of light ellipticity is $\xi\in[-1,1]$, $\xi=\pm 1$ denotes left and right circular polarization. ${\alpha}^{\left(0\right)}\left(\omega  \right), {\alpha}^{\left(1\right)}\left(\omega  \right), {\alpha}^{\left(2\right)}\left(\omega  \right)$ are the scalar, vector, and tensor polarizability, respectively. Scalar shift is spin independent. Vector shift acts like an effective magnetic field to generate the linear Zeeman splitting (light shift proportional to $m_{F}$), which depends on the ellipticity of the light and the intersection angle between the laser beam wave vector and magnetic field quantization axis $\hat{e}_{B}$. So there are two methods to control the vector shift, rotating bias magnetic field and changing light polarization. Tensor part is derived from the linearly polarized light, and acts as an effective dc electric field.

For the first excited state of alkali-metal atoms, the fine structure interaction induces the spectral lines of the D1 ($5^{2}S_{1/2}  \rightarrow5^{2}P_{1/2}$) and D2 ($5^{2}S_{1/2}  \rightarrow5^{2}P_{3/2}$) lines. The coefficients of the scalar, vector and tensor shifts of the ground states $5^{2}S_{1/2}$ of $^{87}$Rb atoms in Eq. \ref{eq:1} are given by
\begin{equation}\label{eq:s2}
	\begin{aligned}
		& { {{\alpha }^{\left( 0 \right)}}\left( \omega  \right)\approx -\frac{\pi {{c}^{2}}{{\Gamma }_{{{D2}}}}}{2\omega ^{3}_{0}}\left(\frac{2}{{{\delta }_{{{D2}}}}}+\frac{1}{{{\delta }_{{{D1}}}}} \right)}, \\
		& {{\alpha }^{\left( 1 \right)}}\left( \omega  \right)\approx -\frac{\pi {{c}^{2}}{{\Gamma }_{{{D2}}}}}{2\omega ^{3}_{0}}\left(\frac{1}{{{\delta }_{{{D2}}}}}-\frac{1}{{{\delta }_{{{D1}}}}} \right){g}_{F}{F},\\
		& {{\alpha }^{\left( 2 \right)}}\left( \omega  \right)\approx 0,\\
	\end{aligned}
\end{equation}
where ${{\Gamma }_{{{D}_{2}}}}$ is the decay rate of the excited state for $D_{2}$ line, $\omega _{0}=\frac{1}{3}\omega _{D1}+\frac{2}{3}\omega _{D2}$ is the effective frequency, ${{\delta }_{{{D}_{1}}}}={{\omega }_{~}}-{{\omega }_{D1}}$, ${{\delta }_{{{D}_{2}}}}={{\omega }_{~}}-{{\omega }_{D2}}$ is the frequency detuning of the laser. Therefore, according to Eq. \ref{eq:s2} we only consider the scalar and vector shift in this work. We employ tune-out wavelength for spin-dependent optical lattice, in which ac Stark shifts cancel. Two internal spin states have different tune-out wavelengths when the contributions of both the scalar and vector shifts are included  ~\cite{Wen2021}.

We choose two ground hyperfine Zeeman states of $^{87}$Rb $\left\vert F=2,m_{F}=0 \right\rangle$ of the $F=2$, and $\left\vert 1,1\right\rangle$ of the $F=1$ hyperfine manifold as the two internal spin states. A bias magnetic field with 10~Gauss is applied along the $45^{\circ}$ diagonal line of the square lattice $V_{2}$. We scan the wavelength of the optical lattice beams to determine the tune-out wavelength precisely, as shown in Fig.~S2. The tune-out wavelength for $\left\vert 1,1\right\rangle$ state is determined at 788.28~nm with $\sigma_{-}$ circular polarization as shown in Fig.~S2(c), which balances the contribution of the scalar and vector shift. Thus we choose this wavelength for the lattice $V_{2}$. Note that the tune-out wavelength for $\left\vert 1,1\right\rangle$ state is sensitive to the intersection angle between the laser beam wave vector and magnetic field quantization axis, which requires a careful alignment of the bias magnetic field carefully. The spin state $\left\vert 2,0\right\rangle$ only experiences the square lattice $V_{2}$ with the red-detuning AC stark shift (which is only from scalar shift) as shown in Fig.~S2(d) and (f), in contrast, the spin state $\left\vert 1,1\right\rangle$ experiences no shift.

On the other hand, there is only the contribution of the scalar shift for the spin state $\left\vert 2,0\right\rangle$, the tune-out wavelength for $\left\vert 2,0\right\rangle$ state is 790.02~nm as shown in Fig.~S2(a), which is well known and studied experimentally~\cite{arora2011tune,Wen2021}. We choose this tune-out wavelength of 790.02~nm with $\sigma_{+}$ circular polarization as the wavelength of the lattice $V_{1}$. Thus the spin state $\left\vert 1,1\right\rangle$ experiences the square lattice $V_{1}$ with the red-detuned AC stark shift. In contrast, the spin state $\left\vert 2,0\right\rangle$ see zero light shift. Note that the tune-out wavelength for $\left\vert 2,0\right\rangle$ state is insensitive to the intersection angle between the laser beam wave vector and magnetic field quantization axis. The spin state $\left\vert 1,1\right\rangle$, however, has a different lattice depth in two orthogonal directions of the lattice $V_{1}$ respectively, and feels the lattice $V_{1}$ with the red-detuning AC stark shift (which is only from vector shift at the wavelength of 790.02~nm) as shown in Fig.~S2(b) and (e).

Moir$\acute{\mathrm{e}}$ superlattice can be generated by a small difference in lattice constant or orientation. Since two different wavelengths are used for twisted-bilayer lattices in this work, there is a large-period superlattice with $\Delta\lambda=179$ $\mu$m, much larger than the size of atomic cloud. Therefore, we can adjust the retroreflected concave mirror to load atoms into the lower potential well of the long-period superlattice and neglect the influence on the measurement of moir$\acute{\mathrm{e}}$ pattern. In the future, we can correct this effect of two different wavelengths by using a slight angle lattice beam for $V_{2}$ to ensure the same lattice constant for two lattice potentials.

\noindent \textbf{III. Band structures and flat band}

\begin{figure}[tb]
\includegraphics[width=0.48\textwidth]{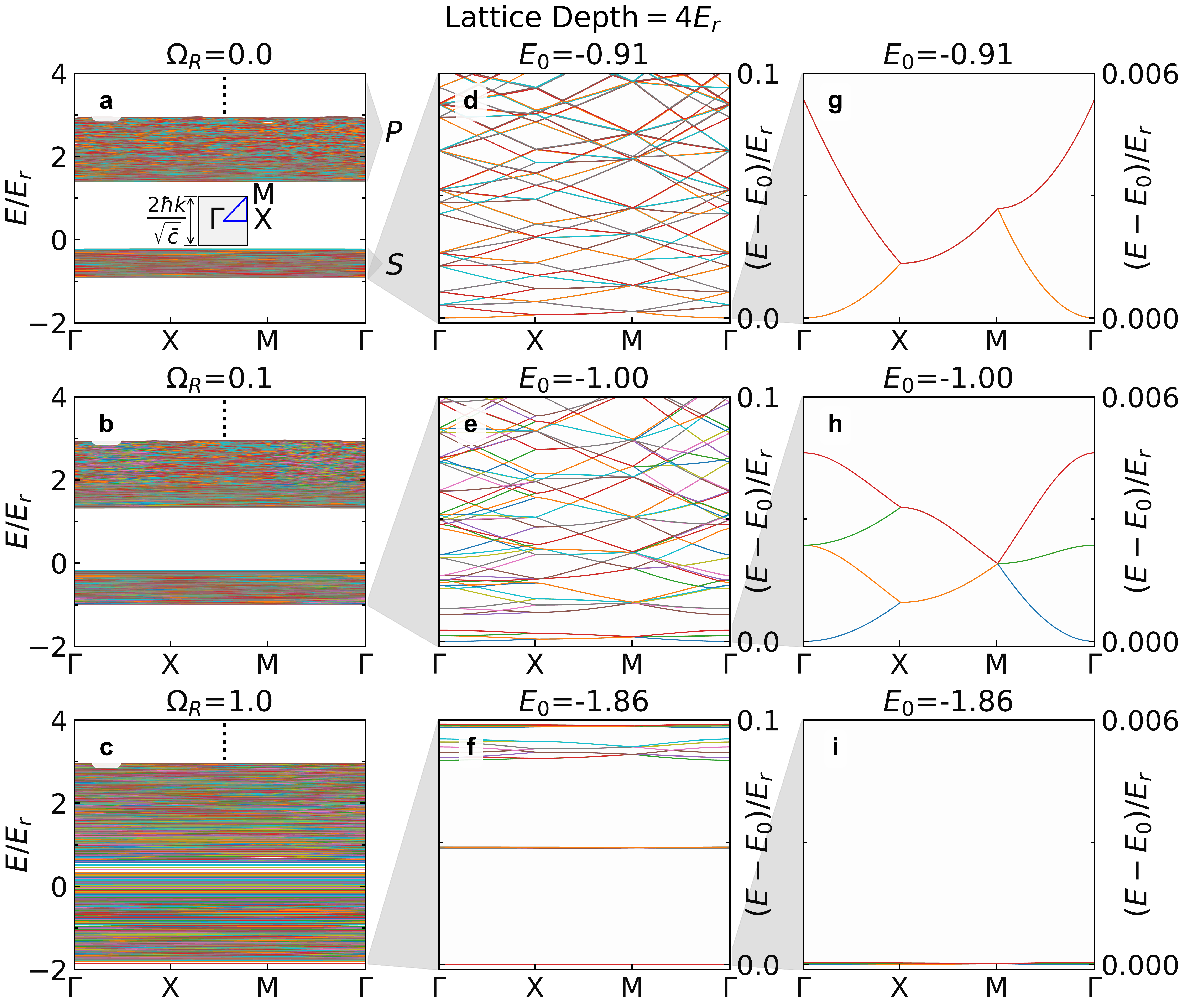}
\setcounter{figure}{2}
\renewcommand{\thefigure}{S\arabic{figure}}
\caption{\textbf{Band structure of the twisted-bilayer optical lattices.} The twist angle of the commensurate optical lattice is $\theta=2\arctan(1/22)$, whose band structure is regarded as an approximation of the experimental case $\theta=5.21^{\circ}$. $\bf{a}$, $\bf{b}$ and $\bf{c}$ show the band structures for the interlayer coupling strength $\Omega_{\mathrm{R}}=0E_{r}$, 0.1$E_{r}$ and 1$E_{r}$ respectively. $\bf{a}$ also gives the band structure without the interlayer coupling in the form of the superlattice minibands within the same reduced Brillouin zone. $\bf{d}$, $\bf{e}$ and $\bf{f}$ are the enlargement of the lowest bands of $\bf{a}$, $\bf{b}$, and $\bf{c}$, respectively. $\bf{g}$, $\bf{h}$ and $\bf{i}$ are the further enlargement of the lowest bands of $\bf{d}$, $\bf{e}$ and $\bf{f}$, respectively. Here, the MW detuning is $\Delta=0$ and $E_{0}$ corresponds to the energy of the lowest band. }
\label{figS3}
\end{figure}

\noindent For square lattices, the commensurate angles $\theta$ satisfy $\tan(\theta/2)=\bar{m}/\bar{n}$, where $\bar{m}$ and $\bar{n}$ are integers. An equivalent condition is $\cos\theta=\bar{a}/\bar{c}$ and $\sin\theta=\bar{b}/\bar{c}$, which can be defined by Pythagorean triples ($\bar{a}^{2}+\bar{b}^{2}=\bar{c}^{2}$, where $(\bar{a},\bar{b},\bar{c})\in N$ are positive integers)~\cite{Wang2020}. The relationship between ($\bar{m}$,$\bar{n}$) and $(\bar{a},\bar{b},\bar{c})$ is $(\bar{m}+i\bar{n})^{2}=(\bar{a}+i\bar{b})$ when $(\bar{m}+\bar{n})\in odd $ and $(\bar{m}+i\bar{n})^{2}=2(\bar{a}+i\bar{b})$ when $(\bar{m}+\bar{n})\in even$. For the commensurate optical lattice, the period of its supercell is given by $\lambda_{sc}=\bar{m}a/\sin(\theta/2)=2\bar{m}\lambda_{mo}$ when $(\bar{m}+\bar{n})\in odd $ and $\lambda_{sc}=\bar{m}a/\sin(\theta/2)=\sqrt{2}\bar{m}\lambda_{mo}$ when $(\bar{m}+\bar{n})\in even $.

Here, we choose the band structure of the commensurate optical lattice with the commensurate angle $\theta=2\arctan(1/22)$ as an approximation of the experimental case. If getting a better approximation of band structure for the experimental case, we can choose the larger supercell to calculate the energy band structure, whose commensurate angle is closer to the experimental case. The band structure $E(\bf k)$ of the commensurate optical lattice can be obtained by solving the stationary Schr{\"o}dinger equation, $H\Psi=E\Psi$, with the Bloch function, $\Psi({\bf r})=\exp^{i {\bf k}\cdot {\bf r}}u(\bf r)$. Here the Hamiltonian $H$ is given as
\begin{equation}\label{eq:2}
H=\left(
\begin{array}{cc}
 -\frac{\hbar^2}{2m}\nabla^2+V_1+\frac{\Delta}{2}  & \Omega_\mathrm{R}  \\
 \Omega_\mathrm{R}  & -\frac{\hbar^2}{2m}\nabla^2+V_2-\frac{\Delta}{2}  \\
\end{array}
\right),
\end{equation}
$u(\bf r)$ is a periodic function with the same periodicity as the coupled lattice. The spin-dependent square optical lattice with a twist angle $\theta$ can be described by the potentials
\begin{subequations}
\label{OLpotential}
\begin{align}
V_1=V_{0}\Big[\sin^2\big(kx\cos\frac{\theta}{2}-ky\sin\frac{\theta}{2}\big)\notag \\
+\sin^2\big(ky\cos\frac{\theta}{2}+kx\sin\frac{\theta}{2}\big)\Big], \\
V_2=V_{0}\Big[\sin^2\big(kx\cos\frac{\theta}{2}+ky\sin\frac{\theta}{2}\big)\notag \\
+\sin^2\big(ky\cos\frac{\theta}{2}-kx\sin\frac{\theta}{2}\big)\Big],
\end{align}
\end{subequations}
where $k=2\pi/\lambda$ is the wave number of lasers for the lattice and $V_0$ describes the lattice depth.
In numerics we first discretize the unit supercell of area $\sqrt{\bar{c}}a\times \sqrt{\bar{c}}a$ in real space ($\bar{c}$ is the largest value in the Pythagorean triple) into $l\times l$ grids, and then diagonalize the effective Hamiltonian for $u(\bf r)$. As shown in Fig.~S3, the band structure approaches the flat band when increasing the interlayer coupling.

Since our system allows for flexible control of the interlayer couplings, the flatband in the lowest energy band can be realized. The Hamiltonian Eq. \ref{eq:2} can be formally diagonalized as
\begin{equation}\label{diag-H}
H=\left(
\begin{array}{cc}
 H_\text{eff}^{+}  & 0  \\
 0  & H_\text{eff}^{-}  \\
\end{array}
\right),
\end{equation}
where
\begin{equation}\label{eff0-H}
H_\text{eff}^{\pm}=h_0\pm h_1,
\end{equation}
with $h_0=-\frac{\hbar^2}{2m}\nabla^2+\frac{V_1+V_2}{2}$, and $h_1=\sqrt{\Omega_\mathrm{R}^2+\frac{(V_1-V_2+\Delta)^2}{4}}$.
In the large interlayer coupling limit, $\Omega_\mathrm{R}\gg V_{0}, \Delta$, the low-energy band structure is encoded in the effective Hamiltonian $H_\text{eff}^{-}$ in the lower-right block, which can be further approximated as
\begin{equation}\label{eff1-H}
H_\text{eff}^{-}\approx -\frac{\hbar^2}{2m}\nabla^2+\frac{V_1+V_2}{2}-\frac{(V_1-V_2+\Delta)^2}{8\Omega_\mathrm{R}}-\Omega_\mathrm{R},
\end{equation}
or in a more rough way
\begin{equation}\label{eff2-H}
H_\text{eff}^{-}\approx -\frac{\hbar^2}{2m}\nabla^2+\frac{V_1+V_2}{2}-\Omega_\mathrm{R},
\end{equation}
The approximated effective Hamiltonians correspond to some effective lattices for a single-layer (single-component) system, separately, $V=\frac{V_1+V_2}{2}-\frac{(V_1-V_2+\Delta)^2}{8\Omega_\mathrm{R}}$ for Eq. \ref{eff1-H}, and $V=\frac{V_1+V_2}{2}$ for Eq. \ref{eff2-H}, with certain global energy shift.
Specifically Eq. \ref{eff2-H} implies that the system becomes the single-layer (single-component) experiencing a twisted optical lattice.

When increasing the interlayer coupling into the strong region, the long-wavelength moir$\acute{\mathrm{e}}$ potential becomes deeper, so atoms in the lowest band are isolated at a larger spatial scale (moir$\acute{\mathrm{e}}$ wavelength), which enhances the wavefunction localization and contributes to the creating of flatband. The single-layer system with a twisted optical lattice (approximation at the strong interlayer coupling limit) admits a flatband structure in the lowest band, which has been studied experimentally in photonic system~\cite{Wang2020,Huang2016,Fu2020}. The moir$\acute{\mathrm{e}}$ flat bands have several advantages. First, the flat bands quench kinetic energy scales (wavefunction localization), thereby drastically enhance the role of interactions and amplify the effects of interactions. Second, the moir$\acute{\mathrm{e}}$ superlattice leads to the emergence of minibands within a reduced Brillouin zone. The small Brillouin zone means that low atomic densities are sufficient for full filling or depletion of the superlattice bands, which is easily controlled in experiment.

\noindent \textbf{IV. Theoretical calculation of the modified superfluid to insulator transition}

\noindent In the mean-field approximation, the system for the superfluid phase can be well described by the coupled Gross-Pitaevskii (GP) equations
\begin{eqnarray}
\label{GPequation}
%\begin{align}
%i\hbar\frac{\partial \psi_{1}}{\partial t} =[-\frac{\hbar^2}{2m}\nabla^2+\frac{1}{2}m\omega^2_{\perp}(x^2+y^2)+V_1+\notag \\
%\eta g_{11}|\psi_1|^2+\eta g_{12}|\psi_2|^2 ]\psi_1+\hbar\Omega_\mathrm{R}\psi_2, \\
%i\hbar\frac{\partial \psi_{2}}{\partial t} =[-\frac{\hbar^2}{2m}\nabla^2+\frac{1}{2}m\omega^2_{\perp}(x^2+y^2)+V_2+\notag \\ \eta g_{12}|\psi_1|^2+\eta g_{22}|\psi_2|^2  ]\psi_2+\hbar\Omega_\mathrm{R}\psi_1,
i\hbar\frac{\partial \psi_{1}}{\partial t} =\Big[-\frac{\hbar^2}{2m}\nabla^2+\frac{1}{2}m\omega^2_{\perp}(x^2+y^2)+V_1+\notag \\
\eta g_{11}|\psi_1|^2+\eta g_{12}|\psi_2|^2 \Big]\psi_1+\hbar\Omega_\mathrm{R}\psi_2, \\
i\hbar\frac{\partial \psi_{2}}{\partial t} =\Big[-\frac{\hbar^2}{2m}\nabla^2+\frac{1}{2}m\omega^2_{\perp}(x^2+y^2)+V_2+\notag \\ \eta g_{12}|\psi_1|^2+\eta g_{22}|\psi_2|^2 \Big]\psi_2+\hbar\Omega_\mathrm{R}\psi_1,
%\end{align}
\end{eqnarray}
where the MW detuning is $\Delta=0$ and the wave function is normalized as $\sum_i\int |\psi_i|^2 d\mathbf{r}=N$, with $N$ the total atom number. The strong confinement along the $z$ axis gives rise to the quasi-2D interaction strengths represented by a reduction coefficient $\eta$ multiplied by $g_{ij}=4\pi\hbar^2a_{ij}/m$, where $\eta^{-1}=\sqrt{h/m\omega_z}$ defines the characteristic length along the z axis, and $a_{ij}$ is the 3D s-wave scattering length. In the experiment, the trapping frequency $\omega_z\approx 2\pi\times 1\mathrm{kHz}$, and the scattering length for the $^{\mathrm{87}}$Rb atoms is about $a_{ij}\approx 100a_B$ with $a_B$ the Bohr radius. This implies that even though the system is thermodynamically 2D, the collisions still keep their 3D character with $\eta^{-1}\gg a_{ij}$. Considering the similarity in scattering lengths $a_{11}$, $a_{22}$ and $a_{12}$ for the $^{\mathrm{87}}$Rb atoms, in the calculation we focus on the SU(2) symmetric interaction with $g=g_{11}=g_{22}=g_{12}$. In addition to the intercomponent atomic interaction, the two components are also coupled by a microwave pulse, which causes Rabi oscillations with frequency $\Omega_\mathrm{R}$.

By using the imaginary time evolution method, one can solve the GP equations numerically for the ground states in the harmonic trap. Theoretically the non-commensurate twist angle $\theta=5.21^{\circ}$ should be a localized single particle ground state while the commensurate angle $\theta=2\arctan{\frac{1}{22}}$ gives rise to extended ground states in the absence of interactions. Experimentally the interatomic interaction is dominant, and always leads to extended many-body states with the aperiodic and periodic bilayer lattices becoming almost indistinguishable.

The phase transition from superfluid to Mott-insulator can be well described by the Bose-Hubbard model in the tight-binding approximation. For simplicity we consider the interlayer coupling as a quasi-periodically perturbed potential, which leads to a site-dependent energy offset
%\begin{equation}
%M_{i}=M_{R}[\sin^2(i_x\pi\cos\theta+i_y\pi\sin\theta)+\sin^2(i_y\pi\cos\theta-i_x\pi\sin\theta)],
%\end{equation}
\begin{eqnarray}
M_{i}=M_{R}\big[\sin^2(i_x\pi\cos\theta+i_y\pi\sin\theta)\notag \\
+\sin^2(i_y\pi\cos\theta-i_x\pi\sin\theta)\big],\label{Hami28}
\end{eqnarray}
where the subindex $i_x$ and $i_y$ label the position of the $i$-th site in the two-dimensional space. The tight-binding Hamiltonian for one layer then is given by
\begin{equation}
H=-t\sum_{\langle i,j\rangle}b^{\dagger}_i b_j+\frac{U}{2}\sum_i\hat{n}_i(\hat{n}_i-1)+\sum_{i} (M_{i}-\mu)\hat{n}_i,
\end{equation}
where the first term describes the nearest-neighbor tunnelling with $b^{\dagger}$ and $b$ being the creation and annihilation operators, and the second term represents the on-site interaction. The hopping amplitude $t$ is considered to be site-independent for weak interlayer coupling and can be estimated by $t=\frac{4}{\sqrt{\pi}}E_r\left(V_0/E_r\right)^{3/4}e^{-2\left(V_0/E_r\right)^{1/2}}$. The local repulsion $U$ depends on the depth of the optical lattice, and is given by $U=\sqrt{8/\pi}ka_sE_r\left(V_0/E_r\right)^{3/4}$~\cite{Zwerger_2003}. The chemical potential $\mu$ controls the average number of atoms in the moir\'{e} lattice.

The mean-field phase diagram (Fig. 5a in the Main text) can be mapped by using the Gutzwiller method, which expands the local state $|\psi_i\rangle$ at site $i$ in the Fock basis~\cite{PhysRevB.45.3137,Sheshadri_1993}. When the interlayer coupling $M_i=0$, the system is reduced to the standard Bose-Hubbard model~\cite{PhysRevB.40.546}, which includes two phases, the superfluid phase and the Mott insulator phase~\cite{PhysRevLett.81.3108,Greiner2002}. While the superfluid phase is identified by the superfluid order parameter $\langle \hat{b}_i\rangle\neq 0$ and an arbitrary filling of the atoms on the site, the Mott insulator phase emerges with an integer number of atoms per site with $\langle \hat{b}_i\rangle=0$. When the interlayer coupling $M_i\neq 0$, the persistent coherence of the moir$\acute{\mathrm{e}}$ and primary lattice length scale, as well as density distribution in real space can be used to distinguish the phases, which is determined by the order parameter $\langle \hat{b}_i\rangle$ and the filling of the atoms on the site $n$ as shown in Fig. S4. The chemical potential $\mu/U=1$ is considered in this calculation.

\begin{figure*}[tb]
\includegraphics[width=0.95\textwidth]{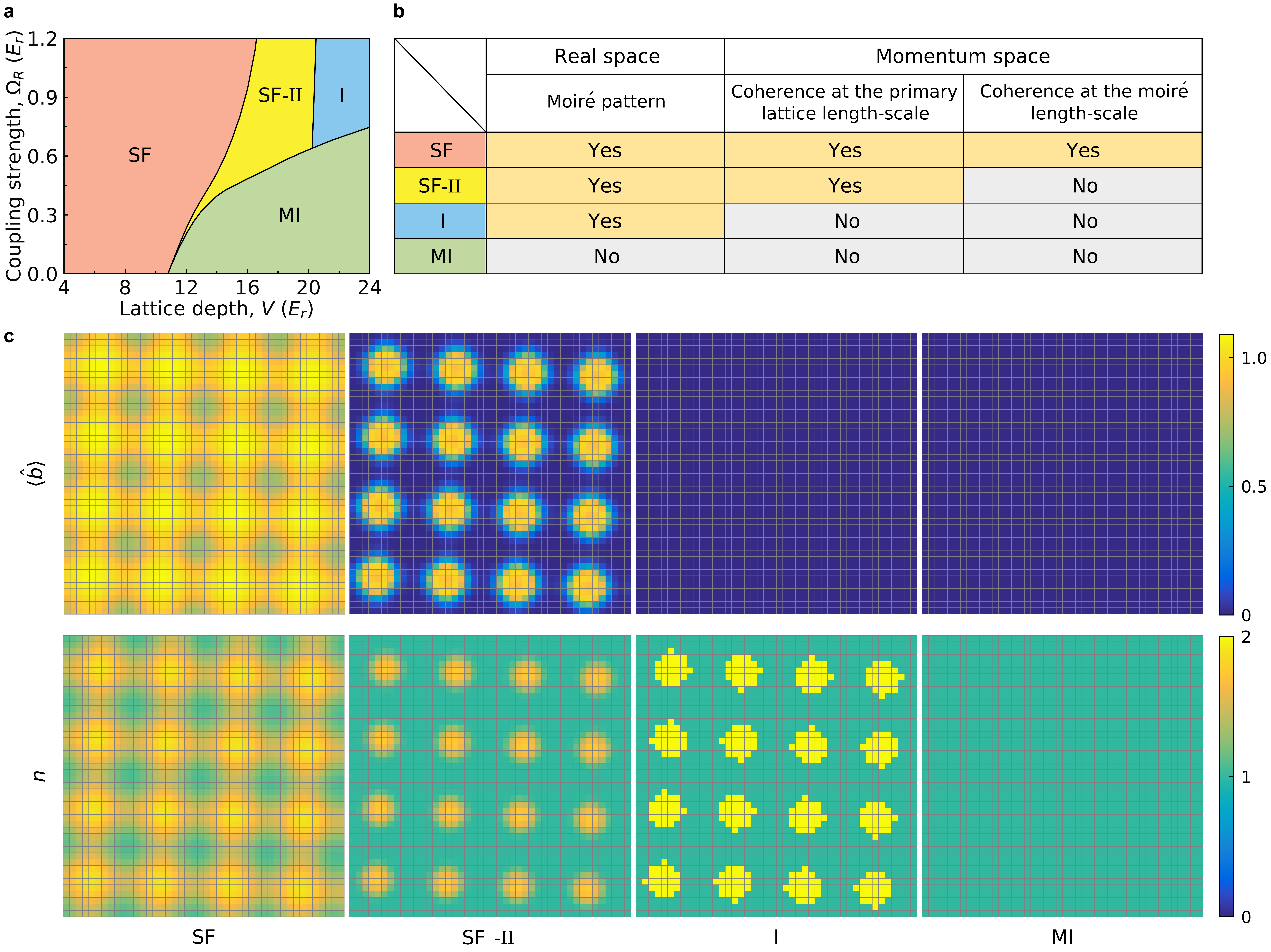}
\setcounter{figure}{3}
\renewcommand{\thefigure}{S\arabic{figure}}
\caption{ \textbf{Characteristics of the different phases.}
$\bf{a}$ Phase diagram, where SF, SF-II, MI, and I refer to superfluid, superfluid only with short-range coherence, Mott insulator, and insulator. $\bf{b}$ Table shows the features of the different phases. $\bf{c}$ Plots of the order parameter $\langle \hat{b}_i\rangle$ and the filling of the atoms on the site $n$ for the different phases. Parameters ($V/E_r$, $\Omega_\mathrm{R}/E_r)$ are (10,0.6), (15,0.6), (23,0.3) and (23,1.1) for the plots from left to right respectively. The chemical potential $\mu/U=1$ is considered.}
\label{sFig1}
\end{figure*}

Phase diagram with zero temperature and the homogeneous system is predicted theoretically as shown in Fig. S4a, in which four phases of superfluid (SF), superfluid II (SF-II), Mott insulator (MI), and insulator (I) are included. The SF-II phase is a state with superfluid domains embedded in a gapped insulate state, which is caused by interlayer coupling. So SF-II phase can be identified by checking the disappearance of the moir$\acute{\mathrm{e}}$-scale long-range correlation with vanishing moir$\acute{\mathrm{e}}$ lattice momentum but the remaining of short-range coherence with residual primary lattice momentum. At the same time, the SF-II phase supports the moir$\acute{\mathrm{e}}$ pattern in the real-space. Since MI is an incompressible insulator for integer filling factor with a gap $U$ for particle–hole excitations induced by the onsite interaction $U$, the moir$\acute{\mathrm{e}}$ pattern in the real-space appears only when the interlayer coupling strength is larger than $U$ to break this gap. Therefore, the I phase supports the moir$\acute{\mathrm{e}}$ pattern in the real-space and no spatial coherence at all scales, which approaches the MI phase without the moir$\acute{\mathrm{e}}$ pattern in the real-space in the limit of weak interlayer coupling. Here, the phase transition between the SF to the MI phase should have an intermediate phase SF-II. Obviously, the coherence is lost almost simultaneously in all length scales at the critical point at very weak interlayer coupling only for zero temperature as shown in Fig. 5a. In contrast, at finite temperature, the thermodynamic quantities behave smoothly near the critical point and the long-range moir$\acute{\mathrm{e}}$ coherence is lost more before the short-range primary lattice length. Therefore, there exists an SF-MI critical regime, which seems likely to be a thermodynamic phase and is not predicted theoretically at zero temperature. The SF-MI critical regime has the same coherence as SF-II without the moir$\acute{\mathrm{e}}$-scale long-range correlation but with the short-range coherence. However, the SF-MI critical regime does not support the moir$\acute{\mathrm{e}}$ pattern in the real-space which is distinguished from SF-II. In fact, the insulator phase presents special characteristics due to the strong interaction and quasi-disorder induced by the larger interlayer coupling, such as similar to Bose glass insulator from the model of disordered strongly interacting bosonic system~\cite{PhysRevB.40.546,PhysRevLett.67.2307,PhysRevB.53.2691}.

\end{document}